%% file: main.tex
\PassOptionsToPackage{hyphens}{url}

\documentclass[runningheads]{llncs}

\usepackage[T1]{fontenc}
\usepackage{graphicx}

\usepackage{amsmath}
\usepackage{amssymb}

\usepackage[table]{xcolor}
\usepackage{booktabs}
\usepackage{array}
\usepackage{multirow}
\usepackage{makecell}
\usepackage{threeparttable}
\usepackage{adjustbox}
\usepackage{siunitx}

\usepackage{calc}
\usepackage{ifthen}
\usepackage{etoolbox}
\usepackage[normalem]{ulem}

\robustify\uline
\robustify\tnote

\usepackage{algorithm2e}

\usepackage{tikz}
\usetikzlibrary{positioning, shapes, arrows}

\usepackage{pgfplots}
\pgfplotsset{compat=1.18}
\usepgfplotslibrary{polar}

\usepackage[final]{microtype}
\usepackage{hyperref}

\hypersetup{
    colorlinks=true,
    urlcolor=blue,
    linkcolor=black,
    citecolor=black
}

\usepackage[abbreviations]{glossaries-extra}
\glsdisablehyper
\setabbreviationstyle[acronym]{long-short}

\NewDocumentCommand{\heatgreen}{O{} m O{}}{%
  \pgfmathsetmacro{\saturation}{0.05 + 0.4 * #2}%
  \pgfmathsetmacro{\brightness}{1 - 0.4 * #2}%
  \xdef\tempcolor{0.33,\saturation,\brightness}%
  \cellcolor[hsb]{\tempcolor}%
  \ifthenelse{\equal{#1}{bold}}{%
    \textbf{%
      #2%
      \ifthenelse{\equal{#3}{}}{}{\scriptsize{$\pm$ #3}}%
    }%
  }{%
    \ifthenelse{\equal{#1}{underline}}{%
      \uline{%
        #2%
        \ifthenelse{\equal{#3}{}}{}{\scriptsize{$\pm$ #3}}%
      }%
    }{%
      #2%
      \ifthenelse{\equal{#3}{}}{}{\scriptsize{$\pm$ #3}}%
    }%
  }%
}

\NewDocumentCommand{\heatorange}{O{} m O{}}{%
  \pgfmathsetmacro{\saturation}{0.05 + 0.4 * #2}%
  \pgfmathsetmacro{\brightness}{1 - 0.4 * #2}%
  \xdef\tempcolor{0.08,\saturation,\brightness}%
  \cellcolor[hsb]{\tempcolor}%
  \ifthenelse{\equal{#1}{bold}}{%
    \textbf{%
      #2%
      \ifthenelse{\equal{#3}{}}{}{\scriptsize{$\pm$ #3}}%
    }%
  }{%
    \ifthenelse{\equal{#1}{underline}}{%
      \uline{%
        #2%
        \ifthenelse{\equal{#3}{}}{}{\scriptsize{$\pm$ #3}}%
      }%
    }{%
      #2%
      \ifthenelse{\equal{#3}{}}{}{\scriptsize{$\pm$ #3}}%
    }%
  }%
}

\NewDocumentCommand{\heatblue}{O{} m O{}}{%
  \pgfmathsetmacro{\saturation}{0.05 + 0.4 * #2}%
  \pgfmathsetmacro{\brightness}{1 - 0.4 * #2}%
  \xdef\tempcolor{0.60,\saturation,\brightness}%
  \cellcolor[hsb]{\tempcolor}%
  \ifthenelse{\equal{#1}{bold}}{%
    \textbf{%
      #2%
      \ifthenelse{\equal{#3}{}}{}{\scriptsize{$\pm$ #3}}%
    }%
  }{%
    \ifthenelse{\equal{#1}{underline}}{%
      \uline{%
        #2%
        \ifthenelse{\equal{#3}{}}{}{\scriptsize{$\pm$ #3}}%
      }%
    }{%
      #2%
      \ifthenelse{\equal{#3}{}}{}{\scriptsize{$\pm$ #3}}%
    }%
  }%
}

\newacronym{mtl}{MTL}{multi-task learning}
\newacronym{stl}{STL}{single-task learning}
\newacronym{cmap}{cmAP}{class-mean average precision}

\newacronym{eat}{EAT}{End-to-end Audio Transformer}
\newacronym{birdmae}{BirdMAE}{Bird-Masked Autoencoder}
\newacronym{protoclr}{ProtoCLR}{Prototypical Contrastive Learning of Representations}

\newacronym{cvt}{CvT}{Convolutional Vision Transformer}
\newacronym{vit}{ViT}{Vision Transformer}

\newacronym{xc}{XC}{Xeno-canto}
\newacronym{xcl}{XCL}{Xeno-canto-Large}

\newacronym{lp}{LP}{linear probing}
\newacronym{ap}{AP}{attentive probing}
\newacronym{ft}{FT}{full fine-tuning}
\newacronym{dwa}{DWA}{dynamic weight averaging}
\newacronym{gradnorm}{GradNorm}{gradient normalisation}

\begin{document}

\title{Adaptive Loss Balancing for Multi-Task Bioacoustic
Classification of Bird Species and Call Types}

\titlerunning{Adaptive Loss Balancing for Bioacoustic Classification}

\author{
Paria Vali Zadeh\inst{1}\orcidID{0009-0007-8396-1585}
\and
Sven Tomforde\inst{1}\orcidID{0000-0002-5825-8915}
}

\authorrunning{P. Vali Zadeh and S. Tomforde}

\institute{
Kiel University, Kiel, Germany\\
\email{paria.vali.zadeh@cs.uni-kiel.de}\\
\url{https://www.uni-kiel.de}
}

\maketitle
\begin{abstract}
%The abstract should briefly summarise the contents of the paper in
%150--250 words.

Reliable analysis of bird vocalisations in passive acoustic monitoring requires models that can handle multiple, imbalanced annotation targets. We extend BirdCallNet for joint species and call-type classification on the long-tailed WiWa forest-soundscape dataset and investigate how task-loss balancing interacts with pretrained representations and adaptation depth. Four bird-domain encoders--ConvNeXt$_{\mathrm{BS}}$, EAT, BirdMAE, and ProtoCLR--are evaluated with separate species and call-type heads under linear probing, attentive probing, and full fine-tuning. A manually tuned fixed objective is compared with homoscedastic uncertainty weighting and Dynamic Weight Averaging across all three adaptation regimes, while GradNorm is evaluated only under full fine-tuning.

The results indicate that the factorised multi-task formulation yields the most consistent improvements over the combined single-task baseline for call-type recognition, while its effect on species recognition depends more strongly on the adaptation regime. Full fine-tuning is not consistently optimal: ConvNeXt$_{\mathrm{BS}}$ achieves the highest mean species performance under linear probing, whereas BirdMAE provides the strongest call-type performance under attentive probing. Adaptive weighting benefits species recognition more consistently than call-type recognition. Uncertainty weighting is particularly effective for species recognition under attentive probing, whereas Dynamic Weight Averaging is generally stronger for the same task under full fine-tuning. GradNorm achieves competitive call-type performance for selected backbones but consistently underperforms the other weighting strategies for species recognition, while incurring higher computational and memory costs. Overall, the preferred loss-balancing strategy depends on the backbone, adaptation regime, and target task, while frozen-backbone adaptation can provide a more favourable performance--efficiency trade-off than end-to-end fine-tuning.

%\keywords{First keyword  \and Second keyword \and Another keyword.}
\end{abstract}

\section{Introduction}
\label{sec:introduction}

Bird populations are increasingly affected by anthropogenic pressures, including habitat loss and fragmentation, land-use change, climate change, and mortality associated with human-made structures \cite{ii_bird_2005}. 
These pressures can alter species distributions, reduce population sizes, and increase the risk of local or global extinction \cite{ii_bird_2005,brunk_assessing_2025}. 
Large-scale studies have reported substantial declines in bird abundance, further emphasising the need for reliable monitoring approaches that can track population changes and support conservation planning \cite{rosenberg_decline_2019,drewitt_collision_2008,loss_birdbuilding_2014}.

Estimating bird population status and understanding the drivers of population change remain challenging. 
Conventional field surveys can provide high-quality observations, but they are often labour-intensive, spatially restricted, and difficult to repeat continuously across large temporal and geographic scales \cite{gibb_emerging_2019,tuia_perspectives_2022}. 
Moreover, estimating anthropogenic mortality or population-level effects is complicated by incomplete observations, detection bias, and variation in survey conditions \cite{loss_birdbuilding_2014,drewitt_collision_2008}. 
These limitations motivate the development of complementary monitoring methods, including passive acoustic monitoring, that can extend ecological surveys across broader spatial and temporal scales \cite{gibb_emerging_2019,ross_passive_2023}.

Bioacoustic monitoring offers a scalable and non-invasive way to observe wildlife and acoustic ecosystems without requiring continuous human presence in the field \cite{gibb_emerging_2019,tuia_perspectives_2022}. 
Passive acoustic monitoring can capture long-term recordings across different habitats and seasons, making it particularly useful for studying vocal animals such as birds \cite{ross_passive_2023,miron_what_2025}. 
Bird vocalisations can provide information not only about species presence, but also about behaviour, communication, reproductive activity, and ecological interactions \cite{marler_natures_2004,teixeira_bioacoustic_2019,lewis_uses_2021}. 
Therefore, acoustic data have the potential to support ecological analyses that go beyond species occurrence, including behavioural and conservation-relevant interpretations of vocal activity \cite{teixeira_bioacoustic_2019,lewis_uses_2021}.

Despite this potential, much of the existing work in automated bird sound analysis focuses primarily on species identification. 
Species-level labels are essential for biodiversity monitoring, but they do not capture the full biological information encoded in bird vocalisations \cite{teixeira_bioacoustic_2019,lewis_uses_2021}. 
Different vocalisation types may reflect distinct behavioural contexts, including mate attraction, territorial defence, alarm, contact, begging, and social interaction \cite{teixeira_bioacoustic_2019,lewis_uses_2021}. 
Consequently, call-type information can help connect acoustic detections to behaviourally and ecologically meaningful events rather than treating all vocal activity as equivalent evidence of species presence \cite{teixeira_bioacoustic_2019,lewis_uses_2021}.

The importance of such information is increasingly apparent in conservation bioacoustics. 
Vocal behaviour can itself be influenced by demographic and social conditions in threatened populations \cite{valderrama_effects_2013,crates_loss_2021}. 
For instance, in the critically endangered regent honeyeater, severe population decline has been linked to erosion of vocal culture, reduced song complexity, and reproductive fitness costs \cite{crates_loss_2021}. 
This suggests that acoustic monitoring can provide conservation-relevant information beyond occurrence alone, particularly when analyses distinguish between different forms of vocal behaviour \cite{crates_loss_2021,teixeira_bioacoustic_2019,lewis_uses_2021}.

Although recent bioacoustic benchmarks have accelerated progress in automated bird sound recognition, most learning tasks are still formulated around taxonomic prediction. 
Large-scale resources such as \gls{xc}~\cite{vellinga_xeno-canto_2015} and BirdSet~\cite{rauch_birdset_2025} provide valuable foundations for species-level modelling, yet they offer limited support for systematically studying vocalisation-type prediction. 
This creates a methodological gap: models are commonly evaluated on whether they can identify the calling species, while their ability to distinguish behaviourally meaningful vocal categories remains less explored.

Addressing this gap requires not only suitable annotations, but also learning strategies that can handle multiple targets at once. 
In a joint species--call-type setting, the two tasks may differ in difficulty, class imbalance, label granularity, and convergence behaviour. 
A fixed combination of task losses can therefore favour one objective over the other, especially when one label space dominates the optimisation signal. 
\gls{mtl} provides a natural framework for sharing acoustic representations between related targets, but its effectiveness depends strongly on how the task-specific losses are balanced during training \cite{kendall_multi-task_2018,liu_end--end_2019}.

In this article, we study adaptive loss balancing for joint bird species and call-type classification on WiWa, a forest soundscape dataset annotated with both taxonomic and vocalisation-type labels. 
We first characterise the WiWa label space, focusing on the distributional properties and imbalance of species, call-type, and combined species--call-type annotations. 
We then compare a fixed weighted-sum objective with three adaptive loss-balancing strategies: homoscedastic uncertainty weighting~\cite{kendall_multi-task_2018}, \gls{dwa}~\cite{liu_end--end_2019}, and \gls{gradnorm}~\cite{chen_gradnorm_2018}. 
These strategies are evaluated across \gls{lp}~\cite{ghani_global_2023}, \gls{ap}~\cite{el-nouby_scalable_2024} with frozen encoders, and \gls{ft}, using several pretrained audio encoders, including ConvNeXt$_{\mathrm{BS}}$~\cite{liu_convnet_2022}, \gls{eat}~\cite{gazneli_end--end_2022}, \gls{birdmae}~\cite{rauch_can_2025}, and \gls{protoclr}~\cite{moummad_domain-invariant_2024}. 
Through this comparison, we examine how loss weighting affects predictive performance and the trade-off between taxonomic and vocalisation-type recognition.

\paragraph{\bf Contributions.}
This paper makes the following contributions.

First, we provide an extended analysis of the WiWa dataset, focusing on the structure and imbalance of its species and call-type label spaces. 
This analysis characterises the dataset as a challenging long-tailed bioacoustic benchmark and clarifies why joint species--call-type classification requires careful modelling and evaluation.

Second, we reformulate joint bird species and call-type classification as an optimisation problem in which the relative weighting of task-specific losses can substantially affect model behaviour. 
Rather than treating \gls{mtl} only as a fixed combination of two classification losses, we analyse how the relative weighting of the two objectives influences both species and call-type performance.

Third, we introduce and evaluate three adaptive loss balancing strategies for this setting: homoscedastic uncertainty weighting, \gls{dwa} and \gls{gradnorm} . 
These methods are compared with a fixed weighted-sum objective to assess whether adaptive weighting can reduce the dependence on manually selected task weights.

Fourth, we evaluate the proposed loss balancing strategies under the same broad encoder and training-regime settings used for the benchmark, including ConvNeXt$_{\mathrm{BS}}$, \gls{eat}, \gls{birdmae}, and \gls{protoclr} under \gls{lp}, \gls{ap} with frozen encoders, and \gls{ft}. 
This provides a controlled comparison of fixed and adaptive task weighting across different representation-learning conditions.

Fifth, we analyse the trade-off between species recognition and call-type classification under different weighting schemes. 
This analysis shows how multi-task optimisation affects both label spaces and clarifies the role of loss balancing in robust and interpretable bioacoustic classification.

\paragraph{\bf{Extension to the Conference Version.}}
\label{sec:extension_over_conference}

This article is an extended and substantially revised version of our conference paper~\cite{vali_zadeh_birdcallnet_2026}. 
The conference version introduced BirdCallNet and WiWa, and presented an initial evaluation of \gls{stl} and fixed-weight \gls{mtl} for joint bird species and call-type classification. 
It also evaluated several pretrained audio encoders and training regimes as part of this initial benchmark. 
In the present article, we retain the biological motivation, the joint classification setting, and the encoder comparison, but substantially extend both the dataset analysis and the multi-task optimisation study.

First, the extended version provides a more detailed characterisation of WiWa, including a deeper analysis of the species and call-type label spaces and their imbalance. 
This additional analysis clarifies the long-tailed structure of the dataset and motivates the need for methods that can handle multiple unevenly distributed prediction targets. 
Second, the methodological extension introduces three adaptive loss-balancing methods: homoscedastic uncertainty weighting, \gls{dwa}, and \gls{gradnorm}. 
These methods are compared with the fixed weighted-sum objective used in the conference version.

The new loss balancing strategies are evaluated under the same broad experimental setting, including ConvNeXt$_{\mathrm{BS}}$, \gls{eat}, \gls{birdmae}, and \gls{protoclr}, as well as \gls{lp}, \gls{ap} with frozen encoders, and \gls{ft}. 
This allows the revised study to analyse how different weighting strategies affect the balance between species recognition and call-type classification under comparable encoder and training conditions. 
Accordingly, the extended version adds new dataset analysis, new methodological content, new experiments on adaptive task weighting, and a revised discussion of the species--call-type trade-off.

\paragraph{\bf Paper Organization.}
The remainder of this article is organised as follows. 
Section~\ref{sec:background} reviews related work on bioacoustic dataset design and label granularity, pretrained audio encoders, transfer and probing strategies, \gls{mtl}, and adaptive loss balancing. 
Section~\ref{sec:methodology} presents the model architecture, multi-task formulation, encoder adaptation settings, and loss-balancing strategies. 
Section~\ref{sec:datasets} describes the WiWa dataset, label construction, and label-distribution analysis. 
Section~\ref{sec:experimental-setup} describes the training protocol, model selection, and evaluation metrics. 
Section~\ref{sec:results-discussion} reports the experimental results for species and call-type classification and discusses the main findings and limitations. 
Section~\ref{sec:conclusion} concludes the article.

\section{Background and Related Work}
\label{sec:background}

Bioacoustic monitoring increasingly relies on deep learning for large-scale automated species recognition in complex acoustic environments~\cite{stowell_computational_2022,kahl_birdnet_2021}. Recent advances have been driven by pretrained audio encoders, which yield transferable representations suited to downstream tasks with limited or imbalanced annotations~\cite{schwinger_foundation_2025,chen_beats_2022,baevski_data2vec_2022}. This study examines such representations under probing and fine-tuning regimes~\cite{ghani_global_2023,schwinger_foundation_2025}, extended to a dual-label setting for joint species and call-type classification — a formulation that naturally induces a \gls{mtl} problem~\cite{caruana_multitask_1997,zhang_survey_2022,vali_zadeh_birdcallnet_2026}. As the two objectives may differ in imbalance, acoustic separability, and convergence behaviour, adaptive loss balancing is further considered~\cite{kendall_multi-task_2018,liu_end--end_2019}. This section reviews the main lines of work informing these contributions: bioacoustic dataset design and label granularity, pretrained encoders, transfer strategies, \gls{mtl}, and adaptive loss balancing.

\paragraph{Bioacoustic Datasets and Label Granularity.}

Large-scale bioacoustic resources have become foundational to automated avian sound recognition. These include community archives such as \gls{xc}~\cite{vellinga_xeno-canto_2015}, identification systems such as BirdNET~\cite{kahl_birdnet_2021}, and benchmark frameworks such as BirdSet~\cite{rauch_birdset_2025}. Despite their complementary roles, all share a taxonomic labelling paradigm in which recordings are annotated primarily by species.
Sub-species annotation has nonetheless been explored in specialised corpora. Bird-DB provides phrase-type and song-sequence labels~\cite{arriaga_bird-db_2015}, while NIPS4Bplus supports the comparison between species-level and vocalisation-type labels that span calls, songs, and drumming~\cite{morfi_nips4bplus_2019,bravo_sanchez_bioacoustic_2021}. Yet, a consolidated benchmark literature on call-type classification remains absent.
Fine-grained vocalisation labels are biologically motivated: call types correspond to distinct communicative functions, including attraction of partners, territorial defence, alarm signalling, and social coordination~\cite{marler_natures_2004,lewis_uses_2021}, and in declining populations can further index demographic processes such as social isolation and reduced reproductive fitness~\cite{valderrama_effects_2013,crates_loss_2021}.
Jointly annotating species and call type, therefore, introduces two interrelated label spaces that may diverge in class frequency, acoustic separability, and distributional imbalance. These properties are consequential for evaluation, as performance metrics can yield divergent conclusions across sample- and class-level aggregations~\cite{stowell_computational_2022,kahl_birdnet_2021}, and necessitate explicit treatment of long-tailed distributions and cross-label dependencies.

\paragraph{Pretrained Encoders for Avian Acoustics.}

Pretrained audio encoders provide an effective starting point for bioacoustic classification, particularly when downstream data are limited or imbalanced~\cite{hagiwara_aves_2023,schwinger_foundation_2025}. General-purpose models such as data2vec~\cite{baevski_data2vec_2022} and BEATs~\cite{chen_beats_2022} learn transferable representations through self-supervised masked prediction, while domain-adapted models such as AVES~\cite{hagiwara_aves_2023} apply similar objectives to animal vocalisations. Architecturally, convolutional encoders capture local time-frequency patterns salient to bird calls, as exemplified by BirdNET~\cite{kahl_birdnet_2021}; transformer-based encoders such as AST~\cite{gong_ast_2021} model longer-range context via patch-sequence self-attention; and hybrid models combine both inductive biases~\cite{gazneli_end--end_2022}. Recent benchmarks confirm that backbone choice and adaptation strategy substantially affect downstream performance~\cite{schwinger_foundation_2025}.
Species and call-type prediction draw on overlapping but partially distinct acoustic cues. Species recognition relies on spectral and temporal patterns that generalise across conditions~\cite{kahl_birdnet_2021,schwinger_foundation_2025}, whereas call-type prediction depends more on the duration, temporal organisation, and morphology of individual vocal units~\cite{marler_natures_2004,lewis_uses_2021}. We therefore compare convolutional, transformer-based, and hybrid encoders under a common protocol to assess how different representation families transfer to joint classification.

\paragraph{Training Regimes and Probing.}

Recent bioacoustic work has shifted from purely supervised fine-tuning toward transfer learning and self-supervised pretraining, with evaluation protocols that account for class imbalance~\cite{rauch_birdset_2025,schwinger_foundation_2025}. Three adaptation regimes are commonly distinguished. \gls{lp} freezes the encoder and evaluates feature separability directly; \gls{ft} updates the backbone and improves adaptation under domain shift~\cite{ghani_global_2023,schwinger_foundation_2025}; and \gls{ap} offers an intermediate regime by learning to aggregate frozen patch- or frame-level embeddings before classification, which is particularly relevant for transformer-based encoders~\cite{el-nouby_scalable_2024}. Recent foundation-model comparisons suggest that transformers benefit most from \gls{ap}, while strong convolutional encoders remain competitive with linear heads~\cite{schwinger_foundation_2025}. We therefore compare all three regimes to examine how adaptation strategy affects joint species and call-type classification.

\paragraph{Multi-task Learning in Bioacoustics.}

\gls{mtl} is an inductive transfer paradigm in which related tasks are learned jointly through a shared representation, typically implemented as a shared encoder with task-specific output heads~\cite{caruana_multitask_1997,zhang_survey_2022}. Joint supervision can improve generalisation and eliminate the need for independently trained models per task~\cite{caruana_multitask_1997,kendall_multi-task_2018}. This is particularly relevant to bioacoustics, where a single vocalisation carries information at multiple annotation levels~\cite{bravo_sanchez_bioacoustic_2021}. Accordingly, jointly predicting species and higher-level group categories has been shown to outperform single-task training~\cite{kim_multi-task_2020}, suggesting that auxiliary supervision helps preserve broader acoustic structure while learning fine-grained distinctions.
However, \gls{mtl} effectiveness depends critically on how objectives are balanced. When task losses differ in scale or convergence behaviour, a naive weighted sum risks one objective dominating the shared representation~\cite{kendall_multi-task_2018}. This motivates adaptive loss balancing for bioacoustic tasks with heterogeneous label spaces, discussed in the following section.

\paragraph{Adaptive Loss Balancing.}

\gls{mtl} is commonly formulated as the optimisation of a weighted sum of task-specific losses. 
Although fixed loss weights are straightforward to implement, they require manual tuning and implicitly assume that the relative importance of each task remains constant throughout training. 
This assumption may be limiting when tasks differ in loss scale, convergence behaviour, class imbalance, or predictive uncertainty~\cite{kendall_multi-task_2018,liu_end--end_2019}.

Adaptive loss-balancing methods address this limitation by adjusting task weights during optimisation. 
Homoscedastic uncertainty weighting learns task-dependent weights from estimated uncertainty, reducing the contribution of higher-uncertainty tasks while increasing the relative influence of more certain objectives~\cite{kendall_multi-task_2018}. 
\gls{dwa} adjusts weights based on the relative rate of loss reduction, assigning greater emphasis to tasks that improve more slowly and lower emphasis to tasks that converge more rapidly~\cite{liu_end--end_2019}.

Such methods are particularly relevant for joint species and call-type classification, where the two prediction targets may differ in difficulty, imbalance, and optimisation dynamics. 
Adaptive weighting therefore provides a principled alternative to manually selected loss coefficients and enables a more balanced trade-off between taxonomic recognition and vocalisation-type prediction.

\paragraph{Scope of This Work.}
This work focuses on adaptive loss balancing for joint bird species and call-type classification under imbalanced, dual-label conditions. 
Using the WiWa benchmark, we compare fixed and adaptive task-weighting strategies within a shared \gls{mtl} setup. 
Building on the joint classification setting, the study asks: 
(i) how different loss-weighting strategies affect the balance between species recognition and call-type classification, and 
(ii) whether adaptive weighting can reduce reliance on manually selected task weights.

\section{Methodology}
\label{sec:methodology}

This section details the methodological framework used to investigate adaptive loss balancing for multi-task bird sound classification. 
Specifically, we introduce the model architectures based on four pretrained bird-domain encoders and formulate the species and call-type classification problems under both single-task and multi-task learning settings. 
The section then focuses on the joint training objective, where task-specific losses are dynamically balanced using homoscedastic uncertainty weighting~\cite{kendall_multi-task_2018},  \gls{dwa}~\cite{liu_end--end_2019} and \gls{gradnorm}~\cite{chen_gradnorm_2018}.

\subsection{Model Architectures}
\label{subsec:Model Architectures}
The extended experiments employ the same four pretrained encoder families as the conference version~\cite{vali_zadeh_birdcallnet_2026}, with the backbone configuration kept fixed across all training settings. 
This controlled design enables the effects of task formulation and loss-balancing strategy to be examined independently of architectural variation. 
Accordingly, this section does not propose new model architectures, but specifies how the existing pretrained encoders are incorporated into the downstream single-task and multi-task classification framework. 
The selected backbones span convolutional spectrogram-based modelling, waveform-based transformer modelling, masked-autoencoding spectrogram representations, and supervised contrastive spectrogram-transformer representations. 
All encoders are initialised from BirdSet-associated bird-domain checkpoints~\cite{rauch_birdset_2025}. 
For ConvNeXt$_{BS}$~\cite{rauch_birdset_2025}, \gls{birdmae}~\cite{rauch_can_2025}, and \gls{protoclr}~\cite{moummad_domain-invariant_2024}, we use the pretrained weights released in the DBD-research-group BirdSet repository on Hugging Face\footnote{%
\href{https://huggingface.co/DBD-research-group}
{huggingface.co/DBD-research-group}
(last accessed 2 July 2026).%
}.
The \gls{eat}~\cite{gazneli_end--end_2022} encoder is initialised from the corresponding BirdSet-pretrained run archived on Weights \& Biases\footnote{%
\href{https://wandb.ai/deepbirddetect/birdset/runs/pretrain_eat_5_2024-05-19_105101/files}
{wandb.ai/deepbirddetect/birdset}
(last accessed 2 July 2026).%
}.

\subsubsection{ConvNeXt: A Modern Convolutional
Neural Net}
ConvNeXt~\cite{liu_convnet_2022} is a modern convolutional backbone that re-examines the design of conventional ConvNets in light of recent transformer-based architectures. 
Unlike hybrid architectures that incorporate attention mechanisms, ConvNeXt preserves a fully convolutional formulation while adopting a set of architectural refinements, including hierarchical stage organisation, patch-style downsampling, depthwise convolutions, inverted bottleneck blocks, larger convolutional kernels, GELU activations, and layer normalisation. 
These design choices enhance the scalability and competitiveness of convolutional models while retaining their local and hierarchical inductive biases. 
In this study, ConvNeXt$_{BS}$~\cite{rauch_birdset_2025} refers to the BirdSet-adapted ConvNeXt checkpoint used as a spectrogram-based encoder, where log-mel representations are processed as two-dimensional time--frequency inputs to provide a convolutional reference backbone for species and call-type classification.
\subsubsection{EAT: End-to-End Audio Transformer}

\gls{eat}~\cite{gazneli_end--end_2022} is a lightweight end-to-end audio encoder that learns directly from raw waveform samples rather than from fixed time--frequency representations. 
Its architecture combines a one-dimensional convolutional front-end, which performs temporal compression and local feature extraction, with a Transformer encoder that aggregates frame-level information over longer temporal contexts. 
This hybrid design enables the model to learn discriminative audio embeddings while maintaining relatively low computational complexity.

Beyond its architectural design, \gls{eat} was introduced with waveform-level augmentation strategies, including frequency- and phase-domain mixing, to improve robustness and generalisation in end-to-end audio classification. 
These augmentations are particularly relevant because raw-waveform modelling can retain signal properties, such as phase information, that are commonly reduced or omitted in magnitude-based spectrogram representations. 
In our experiments, \gls{eat} is used as a pretrained bird-domain encoder, followed by task-specific heads for species and vocalisation-type classification.

\subsubsection{BirdMAE: Bird-Masked Autoencoder}

\gls{birdmae}~\cite{rauch_can_2025} is a domain-specialised masked autoencoder for self-supervised representation learning in avian bioacoustics. 
It extends the Audio-MAE paradigm~\cite{huang_masked_2022} from general audio to bird vocalisations by applying masked reconstruction to log-mel filterbank spectrograms represented as local time--frequency patches. 
During pretraining, selected spectrogram regions are masked, and the model learns to reconstruct the missing content from the visible context, encouraging the encoder to capture spectro-temporal structures relevant to bird sounds, such as sparse and harmonic vocalisation patterns.

Unlike the general-purpose Audio-MAE~\cite{huang_masked_2022}, \gls{birdmae} is pretrained on the curated \gls{xcl}-1.7M subset of BirdSet\cite{rauch_birdset_2025} and uses a bird-adapted training recipe, including revised input resolution, masking ratio, decoder configuration, training duration, batch size, and mixup-based pretraining. 
The model employs a \gls{vit}-based encoder~\cite{dosovitskiy_image_2021} and is released in base, large, and huge variants. 
After pretraining, the decoder is discarded, and the encoder is used as a pretrained backbone with task-specific heads for species and vocalisation-type classification.

\subsubsection{ProtoCLR: Prototypical Contrastive Learning of Representations}

\gls{protoclr}~\cite{moummad_domain-invariant_2024} is a supervised prototypical contrastive learning framework for learning bird-sound representations under domain shift. 
It combines a compact spectrogram-based encoder with a prototype-level contrastive objective, aiming to learn embeddings that generalise from focal recordings to passive soundscape recordings. 
The encoder is based on \gls{cvt}-13~\cite{wu_cvt_2021}, a hierarchical two-dimensional Transformer architecture that processes log-mel spectrograms through convolutional token embeddings and convolutional Transformer blocks. 
By incorporating depthwise separable convolutions into the attention projections, the backbone can efficiently model local spatiotemporal patterns, while self-attention captures broader contextual dependencies. 
In the configuration used for \gls{protoclr}, the encoder contains 13 Transformer blocks, has approximately 20M parameters, and produces a 384-dimensional clip-level embedding.

The contrastive objective differs from standard supervised contrastive learning by replacing pairwise comparisons of examples with comparisons to class-level prototypes. 
For each class, a prototype is computed as the centroid of its embeddings, and each sample is encouraged to align with its corresponding class prototype while being separated from the prototypes of other classes. 
This formulation promotes intra-class compactness and inter-class separability, while reducing the computational cost of contrastive pretraining. 
As a result, \gls{protoclr} learns bird-sound representations that are both species-discriminative and more robust to recording-domain variation~\cite{moummad_domain-invariant_2024}.

\subsection{Task Formulation}
\label{subsec:task formulation}

We formulate bird vocalisation analysis as a supervised classification problem involving two related prediction targets: species identification and call-type classification. Given an input audio sample $x$, the model predicts a species label $y_{\mathrm{sp}}$ and, in the multi-task setting, an additional call-type label $y_{\mathrm{ct}}$. The extended version of this work investigates not only whether these two objectives benefit from a shared acoustic representation, but also how their respective losses should be balanced during optimisation. To this end, we evaluate four loss-weighting strategies: a naive weighted-sum objective, uncertainty-based weighting, dynamic weight averaging, and gradient-based adaptive weighting. The following sections describe the single-task baseline, the proposed multi-task architecture, and the corresponding loss-balancing strategies.

\subsubsection{Single-Task Learning Baseline}

The \gls{stl} setup serves as the reference configuration for evaluating the benefit of explicit multi-task modelling. 
In this baseline, species and vocalisation-type annotations are collapsed into a single combined label space, where each class represents a unique \emph{species\#call type} combination. 
The model is trained with a single classification head over this combined vocabulary.

Although each training example contains only one positive combined annotation, labels are encoded as binary vectors to remain compatible with segment-level evaluation, where a test window may contain multiple species--call-type combinations. 
Accordingly, the model is optimised using binary cross-entropy with logits over the combined label space. 
This baseline allows us to assess whether modelling species and vocalisation type with separate task-specific heads provides an advantage over a single combined-label formulation.

\subsubsection{Multi-Task Learning Formulation}
\label{subsubsec: Multi-Task Learning Formulation}

\begin{figure}[!htb]
    \centering
    \includegraphics[width=\linewidth]{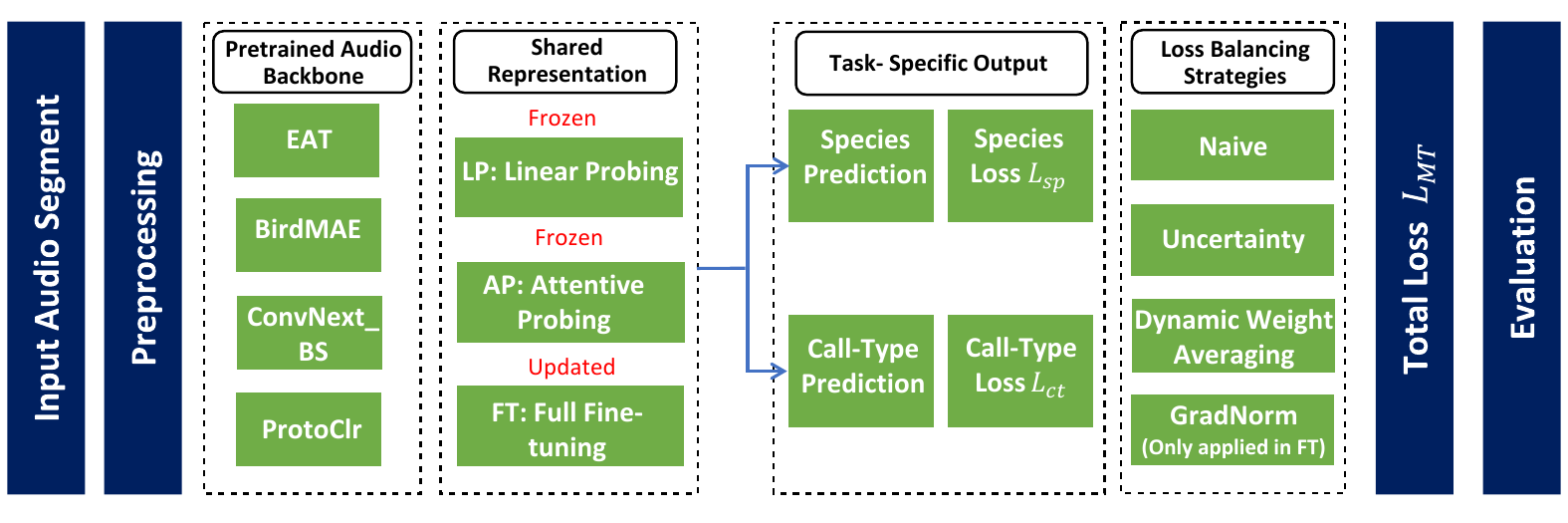}
    \caption{Schematic overview of the proposed multi-task learning (\gls{mtl}) framework for bird vocalisation analysis. Input audio segments are preprocessed and passed through one of the pretrained audio backbones. The shared representation is kept frozen for linear probing (\gls{lp}) and attentive probing (\gls{ap}), whereas it is updated during full fine-tuning (\gls{ft}). Species and call-type predictions are optimised through task-specific losses, which are combined using the evaluated loss-balancing strategies. \gls{gradnorm} is applied only in the \gls{ft} setting because it requires trainable shared parameters.}
    \label{fig:multitask-framework}
\end{figure}

We formulate species identification and vocalisation-type classification as two related prediction tasks learned from a shared audio representation. 
The model follows a hard-parameter-sharing design, in which a common encoder extracts acoustic features and two task-specific heads map the shared representation to separate species and call-type output spaces~\cite{caruana_multitask_1997,zhang_survey_2022}. 
This parallel multi-task configuration allows the model to exploit shared bioacoustic structure while keeping the two label spaces distinct. 
A schematic overview of the architecture is shown in Fig.~\ref{fig:multitask-framework}.

Let $x$ denote an input audio segment. 
The shared encoder $f(\cdot;\theta_s)$ produces a latent representation
\begin{equation}
\mathbf{h} = f(x;\theta_s),
\end{equation}
which is passed to the species and call-type heads:
\begin{equation}
\mathbf{y}_{\mathrm{sp}} = g_{\mathrm{sp}}(\mathbf{h};\theta_{\mathrm{sp}}),
\qquad
\mathbf{y}_{\mathrm{ct}} = g_{\mathrm{ct}}(\mathbf{h};\theta_{\mathrm{ct}}),
\end{equation}
where $\theta_s$ denotes the shared encoder parameters, and $\theta_{\mathrm{sp}}$ and $\theta_{\mathrm{ct}}$ denote the parameters of the task-specific heads. 
The corresponding task losses are denoted by $\mathcal{L}_{\mathrm{sp}}$ and $\mathcal{L}_{\mathrm{ct}}$. 
In our implementation, both losses are computed using binary cross-entropy with logits, since the targets are represented as binary label vectors to support segment-level multi-label evaluation.

\subsubsection{Multi-Task Loss Balancing}
\label{subsubsec:multi-task-loss-balancing}
Since species and call-type prediction may differ in loss scale, convergence behaviour, and task difficulty, we evaluate four loss-weighting strategies for multi-task optimisation: a fixed naive weighted-sum baseline, uncertainty-based weighting~\cite{kendall_multi-task_2018}, \gls{dwa}~\cite{liu_end--end_2019}, and \gls{gradnorm}~\cite{chen_gradnorm_2018}. These methods cover fixed, uncertainty-driven, loss-dynamics-based, and gradient-based balancing mechanisms, respectively. Since \gls{gradnorm} requires gradient norms with respect to shared trainable parameters, it is evaluated only in the \gls{ft} setting; for linear and attentive probing, where the backbone is frozen, we restrict the comparison to loss-based strategies.

\paragraph{Naive Weighted-Sum Loss.}
A common baseline for combining multiple task objectives is to use a naive weighted-sum formulation within the shared-input, hard-parameter-sharing \gls{mtl} setup~\cite{caruana_multitask_1997,zhang_survey_2022}:
\begin{equation}
\mathcal{L}_{\mathrm{naive}}
=
w_{\mathrm{sp}}\mathcal{L}_{\mathrm{sp}}
+
w_{\mathrm{ct}}\mathcal{L}_{\mathrm{ct}} .
\end{equation}
Here, $w_{\mathrm{sp}}$ and $w_{\mathrm{ct}}$ are manually specified task weights that remain fixed throughout training. 
This formulation is simple and transparent, but its performance can be sensitive to the selected weights, particularly when tasks differ in loss scale, difficulty, or convergence behaviour. 
Consequently, the relative task weights must be tuned empirically and cannot adapt to changes in the learning dynamics during optimisation~\cite{kendall_multi-task_2018}.

\paragraph{Uncertainty-Based Loss Weighting.}
To reduce the reliance on manually chosen task weights, we also evaluate uncertainty-based loss weighting~\cite{kendall_multi-task_2018}. 
This method assigns each task a learnable uncertainty parameter and uses it to adapt the effective contribution of the corresponding loss during optimisation. 
For numerical stability, following Kendall et al.~\cite{kendall_multi-task_2018}, we optimise the logarithm of the task variance, denoted by $s_i=\log\sigma_i^2$. 
The objective is defined as
\begin{equation}
\mathcal{L}_{\mathrm{unc}}
=
\sum_{i \in \{\mathrm{sp},\mathrm{ct}\}}
\left(
\exp(-s_i)\mathcal{L}_i
+
\frac{1}{2}s_i
\right).
\end{equation}
This formulation can be interpreted as assigning each task an effective weight proportional to $\exp(-s_i)$, thereby down-weighting tasks with higher estimated uncertainty, while the regularisation term prevents the uncertainty values from increasing without bound. 
The uncertainty parameters are learned jointly with the model parameters.

\paragraph{Dynamic Weight Averaging.}
As a second adaptive strategy, we use \gls{dwa}~\cite{liu_end--end_2019}. 
\gls{dwa} adapts task weights during training by tracking the relative rate of change in each task's loss. 
Inspired by \gls{gradnorm}~\cite{chen_gradnorm_2018}, it aims to balance task learning dynamics over time; however, unlike gradient-based methods, it only requires the numerical task losses and does not need access to internal network gradients. 
This makes \gls{dwa} simple to implement and suitable for our multi-task setup.

For task $i$ at epoch $t$, the relative loss descent rate is defined as
\begin{equation}
\rho_i(t-1)
=
\frac{\mathcal{L}_i(t-1)}
{\mathcal{L}_i(t-2)} .
\end{equation}
The task weight is then computed using a temperature-scaled softmax:
\begin{equation}
w_i(t)
=
\frac{K \exp\left(\rho_i(t-1)/T\right)}
{\sum_{j=1}^{K}\exp\left(\rho_j(t-1)/T\right)} ,
\end{equation}
where $K$ is the number of tasks and $T$ controls the smoothness of the weighting distribution. 
In our setting, $K=2$, corresponding to species and call-type classification. 
The resulting objective is
\begin{equation}
\mathcal{L}_{\mathrm{DWA}}(t)
=
w_{\mathrm{sp}}(t)\mathcal{L}_{\mathrm{sp}}(t)
+
w_{\mathrm{ct}}(t)\mathcal{L}_{\mathrm{ct}}(t).
\end{equation}
Tasks whose losses decrease more slowly receive larger weights, encouraging the model to allocate more optimisation capacity to the currently harder objective.

\paragraph{Gradient Normalisation.}
As a gradient-based adaptive weighting strategy, we also evaluate \gls{gradnorm}~\cite{chen_gradnorm_2018}. 
Unlike \gls{dwa}, which relies only on the evolution of task losses, \gls{gradnorm} adjusts the task weights by directly balancing the gradient magnitudes produced by each task on the shared representation. 
At training step $t$, the weighted multi-task objective is defined as
\begin{equation}
\mathcal{L}_{\mathrm{GN}}(t)
=
\sum_{i \in \{\mathrm{sp},\mathrm{ct}\}}
w_i(t)\mathcal{L}_i(t),
\end{equation}
where $w_i(t)$ denotes the adaptive weight of task $i$. 
For each task, \gls{gradnorm} computes the gradient norm of the weighted loss with respect to a selected set of shared parameters $W$, typically chosen from the shared encoder:
\begin{equation}
G_i(t)
=
\left\|
\nabla_W \left(w_i(t)\mathcal{L}_i(t)\right)
\right\|_2 .
\end{equation}
The method then compares this quantity with a target gradient magnitude that depends on the average gradient norm and on the relative inverse training rate of the task. 
The normalised loss ratio and relative inverse training rate are defined as
\begin{equation}
\tilde{\mathcal{L}}_i(t)
=
\frac{\mathcal{L}_i(t)}{\mathcal{L}_i(0)},
\qquad
r_i(t)
=
\frac{\tilde{\mathcal{L}}_i(t)}
{\frac{1}{2}\sum_{j \in \{\mathrm{sp},\mathrm{ct}\}}\tilde{\mathcal{L}}_j(t)} .
\end{equation}
Tasks with larger $r_i(t)$ are learning more slowly relative to the other task and are therefore encouraged to receive larger gradients. 
\gls{gradnorm} updates the task weights by minimising
\begin{equation}
\mathcal{L}_{\mathrm{grad}}(t)
=
\sum_{i \in \{\mathrm{sp},\mathrm{ct}\}}
\left|
G_i(t)
-
\bar{G}(t)\left[r_i(t)\right]^{\alpha}
\right|_1 ,
\end{equation}
where
\begin{equation}
\bar{G}(t)
=
\frac{1}{2}
\sum_{j \in \{\mathrm{sp},\mathrm{ct}\}}G_j(t)
\end{equation}
is the average task gradient norm, and $\alpha$ controls the strength of the training-rate balancing. 
After each update, the task weights are renormalised so that their sum remains equal to the number of tasks. 
In this way, \gls{gradnorm} reduces the influence of tasks that learn too quickly and increases the contribution of tasks that lag behind, providing an adaptive alternative to manually fixed loss weights.

\section{Dataset}
\label{sec:datasets}
The experiments are conducted on WiWa, a forest-soundscape corpus with annotations at both the species and vocalisation-type levels. A dataset derived from the same underlying recordings is available on Zenodo as the CEB dataset~\cite{martin_ceb_2026}. Rather than treating the dataset solely as an input source, we characterise it as a benchmark whose label-space properties directly influence classification difficulty. We first describe the corpus origin, recording setup, and annotation process, then detail the benchmark construction including label harmonisation, train--test alignment, and the distributional structure of the resulting label spaces.
\subsection{WiWa Corpus}
\label{subsec:WiWa-corpus}
The WiWa corpus originates from a BfN-funded monitoring project examining the effects of wind energy infrastructure on forest bird species in Germany~\cite{reichenbach_auswirkungen_2022}. Recordings were collected between March and June in 2019 and 2020 across 11 forest wind parks in Hesse, Rhineland-Palatinate, and Saarland, using an impact-gradient design in which autonomous recorders were positioned both near wind turbines and at progressively greater distances up to approximately 1~km. Each unit consisted of a Raspberry Pi, external microphone, Sleepy Pi control module, and solar-powered power supply, mounted on tree trunks at approximately 4~m above ground. The campaign produced approximately 5.1~TB of audio (more than 25,000~h), of which 21,090~h entered the original ecological analysis~\cite{reichenbach_auswirkungen_2022}.
Annotation followed an iterative human-in-the-loop protocol. Characteristic vocalisation categories were defined per target species — including song, drumming, flight calls, and alarm calls — and an initial detector trained on manually labelled examples was applied to the full corpus. Expert review and correction of detector predictions were fed back into subsequent training iterations, yielding species- and vocalisation-type annotations under realistic forest soundscape conditions~\cite{reichenbach_auswirkungen_2022}. This dual-label structure directly motivates the joint classification problem examined in this study.
\paragraph{Dataset Composition.}
The training subset comprises 143,981 audio clips of 23~s duration (~920~h, ~273~GB), covering 251 classes: 204 bird classes (100,529 files) and 47 non-bird classes (43,452 files) representing animal and environmental sounds from the same recording sites. Each training instance carries a single candidate class label and, where available, a vocalisation-type label; 14 call-type categories are represented. The test subset consists of 147 passive soundscape recordings totalling 7.35~h (~0.5~GB), with 15,065 annotations across 61 classes and 13 call-type categories. The dataset and its annotations were produced by OekoFor---Ecological Data Collection and Research (GbR)\footnote{%
\href{https://www.oekofor.de/}{oekofor.de}
(last accessed 2 July 2026).%
} and are available to researchers upon request.
\paragraph{Train--Test Asymmetry.}
In avian bioacoustics, focal recordings — centred on a target vocalisation and scalable through community archives such as \gls{xc}~\cite{vellinga_xeno-canto_2015} and the Macaulay Library~\cite{macaulay_library} — are commonly used for training, while passive soundscape recordings, which capture overlapping species and environmental noise simultaneously, serve as the evaluation basis~\cite{rauch_birdset_2025,kahl_birdnet_2021}. This separation reflects a well-documented domain shift: models trained on target-centred data frequently degrade under soundscape conditions~\cite{kahl_birdnet_2021,van_merrienboer_birds_2024}.
WiWa instantiates this asymmetry. The training subset was constructed from BirdNET~\cite{kahl_birdnet_2021} candidate detections validated by experts: each 23~s clip is centred on a 3~s candidate region with approximately 10~s of context on either side, and the annotation specifies candidate class, vocalisation type, BirdNET confidence, and a correctness flag for the 3~s region. Additional species or background sounds may be present in the same clip but are represented as separate instances where validated. The training subset is therefore organised around single validated detections rather than complete multi-label scenes. The test subset, by contrast, consists of passive soundscape recordings in which overlapping vocalisations and environmental noise co-occur naturally, forming an inherently multi-label evaluation problem. This structural asymmetry motivates the benchmark construction described in the following section.
\subsection{WiWa Benchmark Setup}
\label{subsec:benchmark_setup}
We construct the WiWa benchmark by harmonising species and call-type annotations across training and evaluation splits. The evaluation recordings used in the WiWa benchmark correspond to the test partition of the CEB release~\cite{martin_ceb_2026}. The training split serves as the reference label space; labels absent from training are removed from evaluation. Following BirdSet~\cite{rauch_birdset_2025}, we define two evaluation partitions: \emph{test}, containing full soundscape recordings, and \emph{test-5s}, containing non-overlapping 5~s multi-label segments. Segment assignment follows van Merrienboer et al.~\cite{van_merrienboer_birds_2024}: a class is assigned to a 5~s window if its annotation is fully contained within or overlaps it by at least 0.1~s, removing 772 sub-threshold annotation--window overlaps.
Call-type labels are translated and normalised into a common English vocabulary: alarm-related labels are merged into \texttt{alarm call}, \texttt{interaction call} is mapped to \texttt{contact call}, and ambiguous or unsupported labels (\texttt{wingbeat}, \texttt{unknown}, \texttt{something}, \texttt{whistle}, \texttt{vocal}, \texttt{call}) are removed. Background examples are mapped to \texttt{nonbird}. After preprocessing, the training split contains 143,904 samples (including 43,452 background instances), the species label space contains 205 classes (204 bird species and \texttt{nonbird}), and the call-type label space contains 11 classes (10 biological call types and \texttt{nonbird}). The evaluation splits contain 53 species in \emph{test}, 52 in \emph{test-5s}, and 7 biological call types, with 14,961 and 4,868 samples, respectively. 

\begin{figure}[!htbp]
    \centering

    \begin{minipage}[t]{0.74\textwidth}
        \centering
        \includegraphics[width=\textwidth]{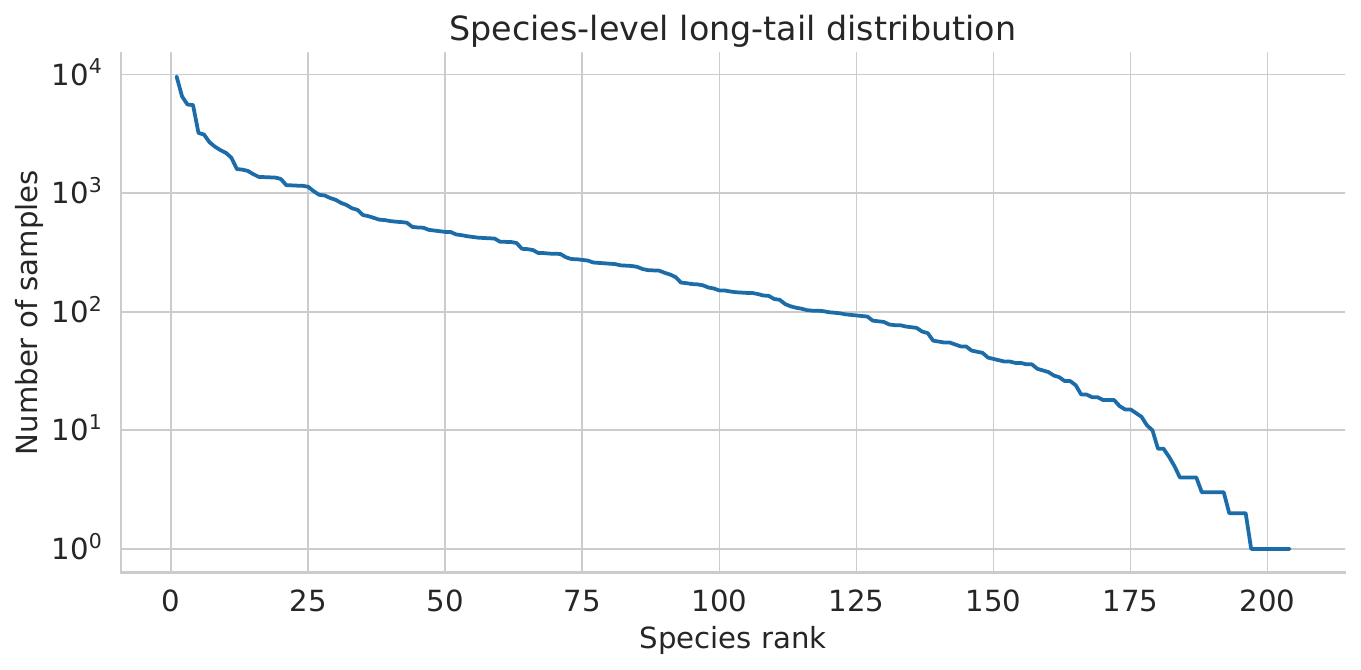}

        \smallskip
        \textbf{(a)} Species-level long-tail distribution.
    \end{minipage}

    \smallskip

    \begin{minipage}[t]{0.74\textwidth}
        \centering
        \includegraphics[width=\textwidth]{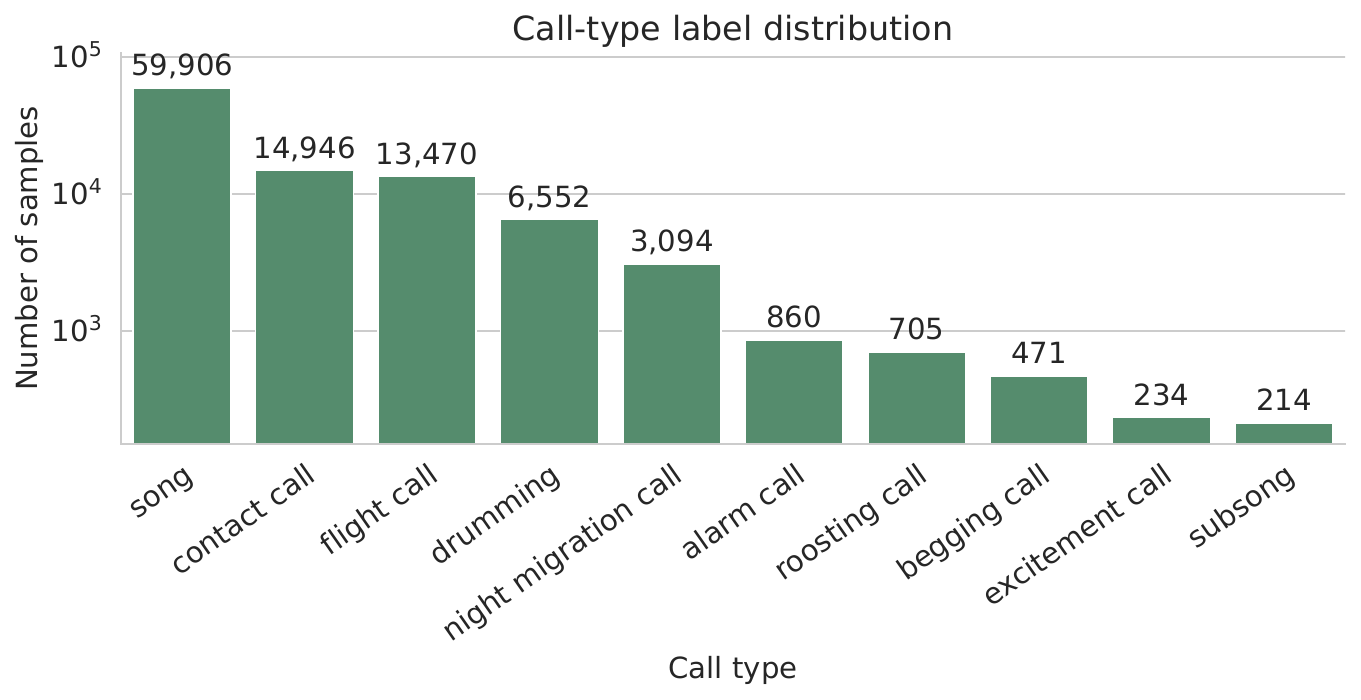}

        \smallskip
        \textbf{(b)} Biological call-type distribution.
    \end{minipage}

    \smallskip

    \begin{minipage}[t]{0.74\textwidth}
        \centering
        \includegraphics[width=\textwidth]{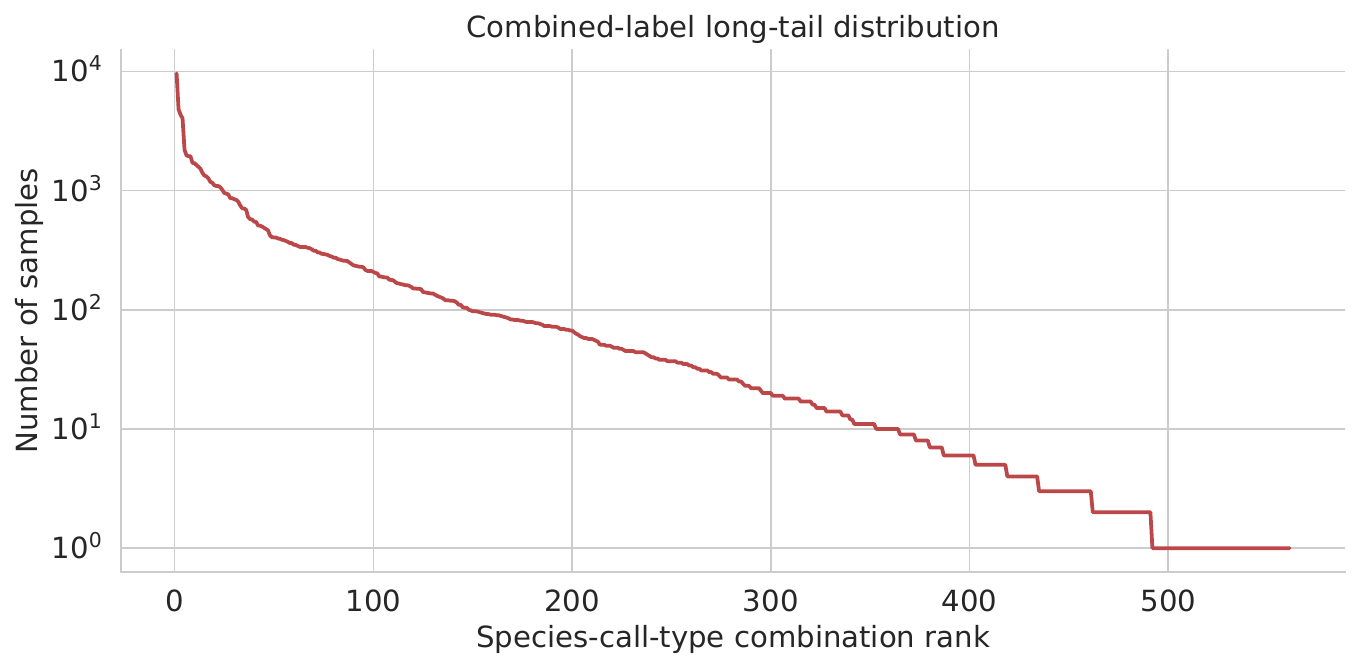}

        \smallskip
        \textbf{(c)} Combined-label long-tail distribution.
    \end{minipage}

    \caption{Label imbalance in the processed WiWa training split:
    (a) species-level long-tail distribution;
    (b) biological call-type distribution on a logarithmic scale; and
    (c) combined-label long-tail distribution over biological
    species--call-type pairs. The \texttt{nonbird} background class
    is excluded from these biological analyses.}
    \label{fig:wiwa-label-imbalance}
\end{figure}

The raw training annotations contain 568 species--call-type combinations. During harmonisation, seven \texttt{wingbeat}-based combinations are removed, leaving 561 biological species--call-type combinations in the processed training split. Including the \texttt{nonbird\#nonbird} background combination gives 562 observed train combinations. Figure~\ref{fig:wiwa-label-imbalance} shows that WiWa is strongly long-tailed across all three label spaces. At the species level, the most frequent class contains 9,632 samples, the median 147, and the rarest a single sample. Call types are dominated by \texttt{song} (59,906 samples), with \texttt{contact call} and \texttt{flight call} distant second and third. The combined label space is more skewed still, with a maximum of 9,631 samples, a median of 26, and several combinations represented by a single example.
\begin{figure}[!htb]
\centering
\includegraphics[width=\textwidth]
{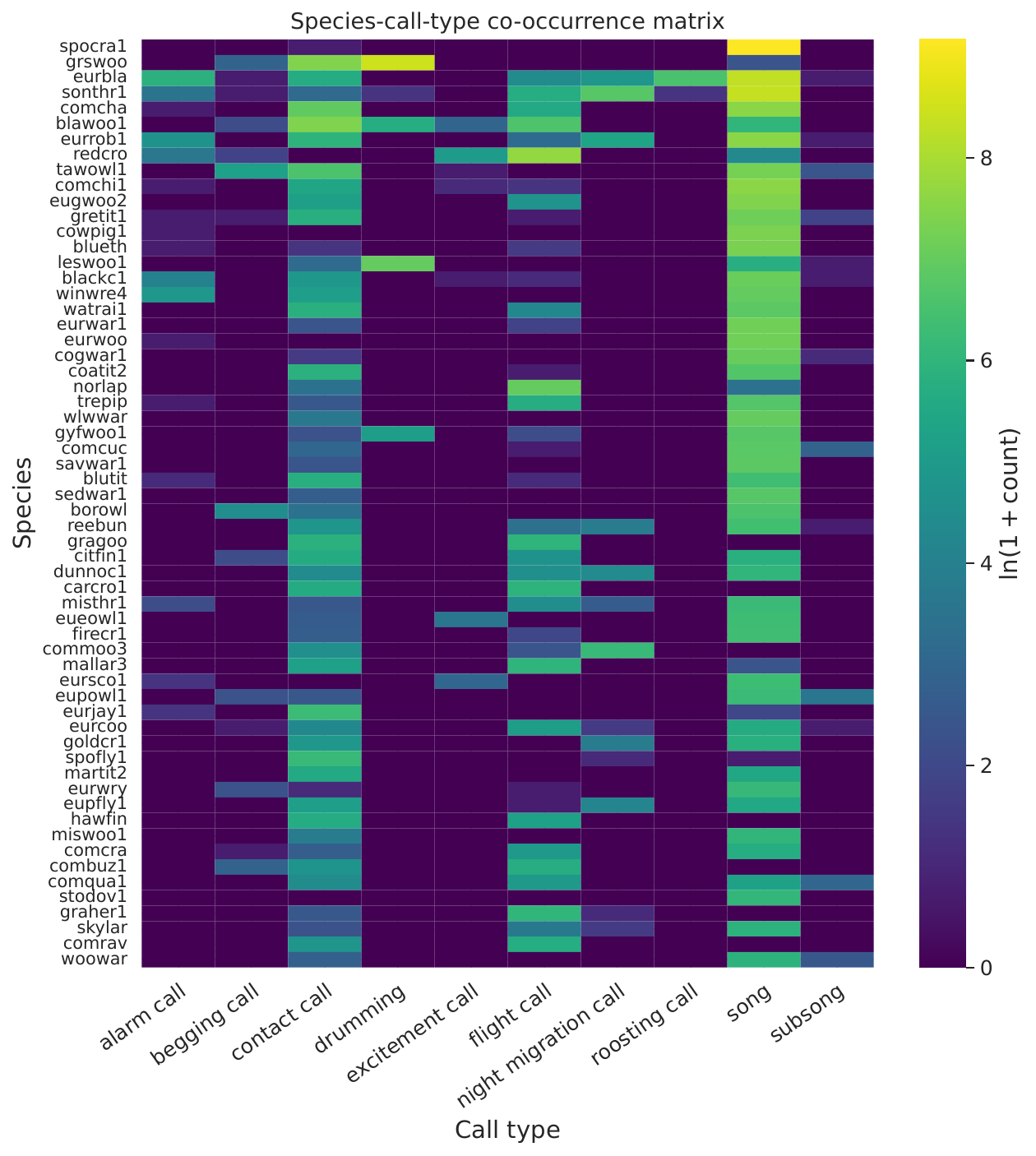}
\caption{Species--call-type co-occurrence matrix for the processed WiWa training split. 
Rows show the 60 biological species with the largest number of samples, and columns show the ten biological call types. 
Cell colours indicate $\ln(1+\mathrm{count})$. 
The \texttt{nonbird} background class is excluded. 
Across the full biological matrix, 561 of 2{,}040 possible species--call-type cells are observed, leaving 72.5\% unobserved.}
\label{fig:wiwa-species-calltype-cooccurrence}
\end{figure}

\begin{figure}[!htb]
\centering
\includegraphics[width=\textwidth]
{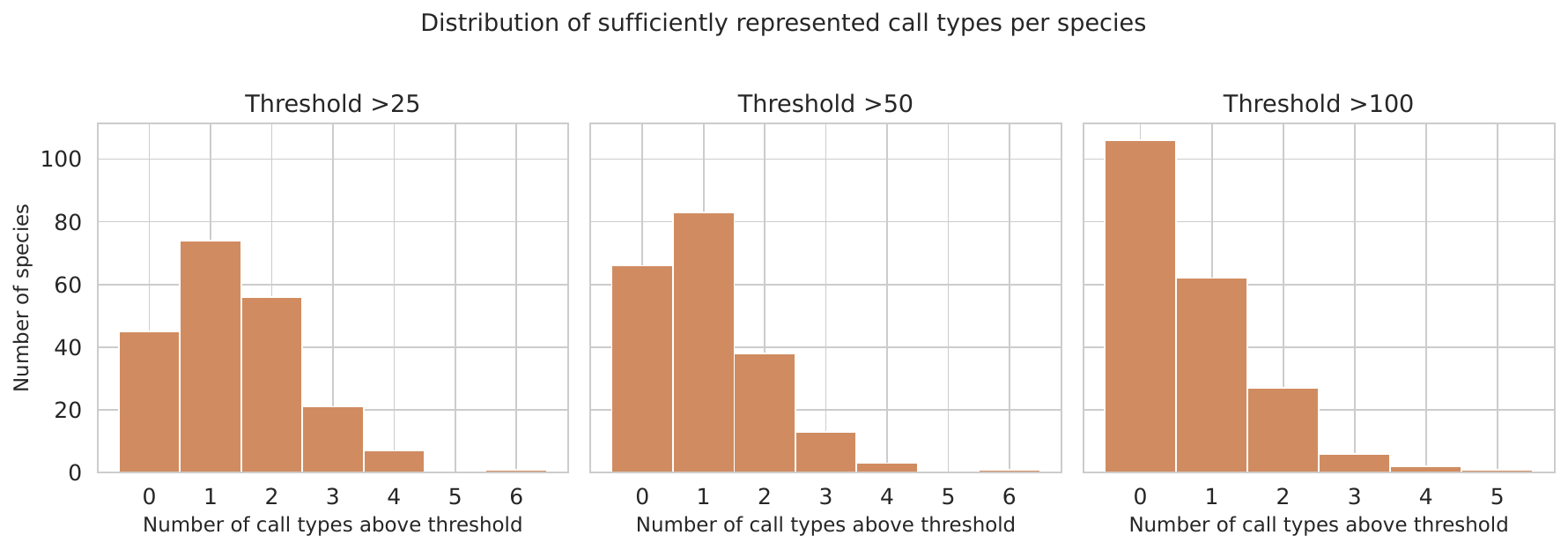}
\caption{Distribution of sufficiently represented biological call types per species in the processed WiWa training split. 
For each species, we count the number of call types with more than 25, 50, and 100 samples. The \texttt{nonbird} background class is excluded.}
\label{fig:wiwa-sufficient-calltypes}
\end{figure}

Beyond marginal imbalance, the biological species--call-type matrix is sparse: 561 of 2{,}040 possible cells are observed, leaving 72.5\% unobserved, as shown in Figure~\ref{fig:wiwa-species-calltype-cooccurrence}. This sparsity is compounded by within-species call-type coverage: only 85, 55, and 36 species have at least two call types represented by more than 25, 50, and 100 samples, respectively (Figure~\ref{fig:wiwa-sufficient-calltypes}). These distributional properties — long-tailed marginals, sparse joint coverage, and uneven within-species call-type representation — define the benchmark's challenge and motivate the comparison between combined-label single-task and multi-task formulations with adaptive loss balancing.

\section{Experimental Setup} \label{sec:experimental-setup}

The experimental framework evaluates adaptive loss-balancing strategies for joint bird species and call-type classification. The design is structured around three primary objectives: (i) assessing the representational advantages of multi-task factorisation over combined single-task formulations; (ii) analysing the trade-offs between species and call-type recognition under fixed versus adaptive weighting; and (iii) evaluating the generalisation of these strategies across diverse encoder architectures and adaptation regimes. To ensure reproducibility and comparability, all configurations are evaluated over three random seeds, and results are reported as the mean across seeds.

\subsection{Task Formulation and Baseline} \label{subsec:task-formulation}
We treat the WiWa benchmark (see Section~\ref{subsec:benchmark_setup}) as a segment-level multi-label classification problem. The model employs two independent linear prediction heads—one for species and one for call types—trained via binary cross-entropy (BCE) with logits. Metrics are computed exclusively over the classes present in the evaluation split to ensure validity. We compare this multi-task approach against a combined single-task baseline \cite{vali_zadeh_birdcallnet_2026}, where species and call-type labels are collapsed into a joint label space (e.g., \textit{species\#calltype}) and predicted via a single head.

\subsection{Architectures and Adaptation Regimes} \label{subsec:adaptation-regimes}

As introduced in Section~\ref{sec:methodology}, we utilise four pretrained bird-domain encoders as shared backbones: ConvNeXt$_{BS}$, \gls{eat}, \gls{birdmae}, and \gls{protoclr}, all initialised from BirdSet checkpoints \cite{rauch_birdset_2025}. To evaluate the interplay between representation quality and loss balancing, we employ three adaptation regimes:

\begin{enumerate} \item \textbf{Linear Probing (LP):} The encoder is frozen; only the linear heads are trained to probe the linear separability of the representations. For an output space with $C$ classes and encoder dimension $d$, the head contains $C(d+1)$ trainable parameters~\cite{alain_understanding_2018,ghani_global_2023}. \item \textbf{Attention Probing (AP):} A trainable multi-head attention pooling module \cite{el-nouby_scalable_2024} is inserted before the heads, allowing content-dependent aggregation while keeping the backbone frozen. For an output space with $C$ classes, the trainable parameter count is $2d^2 + (C+1)d + C$~\cite{el-nouby_scalable_2024} \item \textbf{Full Fine-Tuning (FT):} The entire network is optimised end-to-end, providing maximal capacity for domain-specific adaptation at the cost of higher computational overhead. \end{enumerate}

\subsection{Preprocessing and Augmentation}
\label{subsec:preprocessing}

Preprocessing follows the input requirements of each pretrained backbone. Audio segments are converted to mono, resampled to the corresponding sampling rate, and cropped or zero-padded to the required input duration. \gls{eat} operates directly on 5~s waveform inputs at 22.05~kHz. The remaining encoders use spectrogram-based time--frequency representations: ConvNeXt$_{BS}$ and \gls{birdmae} use 5~s inputs at 32~kHz, whereas \gls{protoclr} uses 6~s inputs at 16~kHz.

During training, we use a BirdSet-style waveform-level augmentation pipeline~\cite{rauch_birdset_2025,schwinger_foundation_2025}. We use multi-label audio mixing with unioned target labels, additive background and coloured noise, and gain perturbation. In addition, call-free VOX segments~\cite{lostanlen_birdvox-dcase-20k_2018} are included as hard-negative examples by assigning all-zero target vectors for both prediction heads. This regularises the models against background-only inputs without introducing spurious positive labels for either task.

\subsection{Training and Optimisation} \label{subsec:training-protocol}

Models are optimised using AdamW \cite{loshchilov_decoupled_2019} with a cosine learning-rate schedule (5\% linear warmup) and gradient clipping ($\ell_2 \leq 0.5$). Training and validation are performed on an 80/20 stratified split. Early stopping is governed by the validation loss, with thresholds adjusted for single-task ($1\times10^{-4}$) versus multi-task ($2\times10^{-4}$) settings to account for the additive loss scale. All configurations are evaluated over three random seeds, and results are reported as the mean across seeds. For the combined single-task baseline and fixed-weight Naive multi-task configurations, two seed runs are retained from our earlier work~\cite{vali_zadeh_birdcallnet_2026} and supplemented with one additional run conducted in the present study. All experiments involving adaptive loss-weighting strategies are newly conducted using three seeds. Hyperparameters per regime are detailed in Table~\ref{tab:training-hyperparams}. The complete experiment configurations and training logs will be made publicly available through Weights \& Biases upon acceptance.

\begin{table}[!htbp]
\centering
\caption{Training hyperparameters per adaptation regime. $^\dagger$Reflects \gls{gradnorm} configurations only due to per-task gradient memory overhead. STL: single-task learning; MTL: multi-task learning.}
\label{tab:training-hyperparams}
\small
\begin{tabular}{lccc}
    \input{Tables/Table1_Hyperparameter}
\end{tabular}
\end{table}

\subsection{Loss-Balancing Strategies} \label{subsec:loss-balancing}

The loss formulations are defined in Section~\ref{subsubsec: Multi-Task Learning Formulation}. Here, we specify the loss-balancing configurations used in the experimental comparison. All configurations are based on the same species and call-type binary cross-entropy losses, but differ in whether task contributions are fixed, learned, dynamically updated from loss trajectories, or adjusted using gradient information. We evaluate four configurations:
\begin{enumerate}
\item \textbf{Fixed weighting (Naive):} A static baseline using the validation-tuned coefficients $\lambda_{\mathrm{sp}}=1.0$ and $\lambda_{\mathrm{ct}}=17.0$ from the conference version of this work~\cite{vali_zadeh_birdcallnet_2026}.

\item \textbf{Uncertainty Weighting:} An adaptive strategy that learns one log-variance parameter per task and uses the estimated homoscedastic uncertainty to adjust the relative task contributions during training~\cite{kendall_multi-task_2018}.

\item \textbf{Dynamic Weight Averaging (\gls{dwa}):} An adaptive strategy that updates task weights according to the relative loss descent rate over a two-epoch window. Following the original study, we set the temperature to $T=2.0$, which was found empirically to perform best across the evaluated architectures~\cite{liu_end--end_2019}.

\item \textbf{Gradient Normalisation (\gls{gradnorm}):} A gradient-based strategy that adapts the task weights by penalising imbalances in task-induced gradient magnitudes. Following the configuration reported in the original study, we set the asymmetry parameter to $\alpha=1.5$~\cite{chen_gradnorm_2018}.
\end{enumerate}

Under \gls{lp} and \gls{ap}, the encoder is frozen, and the shared-gradient signal required by \gls{gradnorm} is unavailable. Consequently, \gls{gradnorm} is evaluated only under \gls{ft}. The resulting strategy--regime assignment is summarised in Table~\ref{tab:loss-balancing-configs}.

\begin{table}[!htbp]
\centering
\caption{Applicability of loss-balancing strategies across adaptation regimes. \checkmark: evaluated; --: methodologically inapplicable (no shared encoder gradients available).}
\label{tab:loss-balancing-configs}
\small
\begin{tabular}{lccc}
    \input{Tables/Table2_lossstrategy_applicability}
\end{tabular}
\end{table}

\subsection{Evaluation Metrics} 
\label{subsec:evaluation}
Final performance is measured on the \textit{test-5s} partition using \gls{cmap}~\cite{kahl_overview_2023}. 
Average Precision is computed independently for each class in a one-vs-rest manner from the corresponding precision--recall curve. 
The class-masked macro-average is then defined as
\begin{equation}
\mathrm{cmAP}
=
\frac{1}{C}
\sum_{c=1}^{C}
\mathrm{AP}(c),
\end{equation}
%where $C$ denotes the number of evaluated classes after applying the class mask, and $\mathrm{AP}(c)$ is the Average Precision for class $c$. 
where $C$ denotes the number of evaluated classes over which the macro-average is computed, and $\mathrm{AP}(c)$ is the Average Precision for class $c$. In the multi-task setting, \gls{cmap} is computed and reported separately for the species and call-type heads. Macro-averaging assigns equal weight to each evaluated class, which is important under the long-tailed label distribution of the benchmark.

\section{Results and Discussion}
\label{sec:results-discussion}

This section evaluates the pretrained backbones across different adaptation regimes and multi-task loss-weighting strategies. Table~\ref{tab:main-results} reports class-masked \gls{cmap} for the combined single-task baseline and the multi-task models under \gls{lp}, \gls{ap}, and \gls{ft}. The reported Score denotes the mean across the Naive, Uncertainty, and \gls{dwa} strategies. Table~\ref{tab:finetune_loss_strategies} extends the \gls{ft} comparison to \gls{gradnorm}. In this study, \gls{gradnorm} is evaluated only under \gls{ft}, where shared backbone parameters are trainable. Its additional task-specific gradient computations required reducing the batch size to 8.

\subsection{Overall Performance}

In the combined single-task setting, \gls{ft} performs best for most backbones, whereas \gls{birdmae} benefits more from \gls{ap}. The multi-task results follow a different pattern: ConvNeXt$_{BS}$ with \gls{lp} achieves the highest mean species performance, while \gls{birdmae} with \gls{ap} provides the highest mean call-type performance.

Compared with the combined-label baseline, the factorised multi-task formulation shows its clearest and most consistent advantage for call-type recognition. For species recognition, the benefit is more dependent on the adaptation regime: performance generally improves under \gls{lp} and \gls{ap}, whereas the \gls{ft} results are mixed and are often lower than the corresponding combined-label baseline. These comparisons should nevertheless be interpreted cautiously because the two formulations operate on different label spaces.

Adaptive loss weighting benefits species recognition more consistently than call-type recognition. For every backbone--adaptation pair, at least one adaptive strategy improves upon Naive for the species task. Adaptive weighting also performs best in most call-type configurations, although the differences are generally smaller. The results further show that \gls{ft} is not uniformly advantageous: frozen-backbone adaptation is consistently preferred for species recognition, while its effectiveness for call types depends on the backbone.

\subsection{Effect of the Adaptation Strategy and Backbone}

The preferred adaptation strategy varies across backbones and prediction tasks. Frozen-backbone adaptation is consistently more effective for species recognition, whereas call-type recognition benefits from \gls{ft} for some, but not all, backbones.

ConvNeXt$_{BS}$ performs best with \gls{lp} for both tasks. This suggests that its pretrained convolutional representation already provides features that can be effectively separated using lightweight prediction heads. Neither \gls{ap} nor \gls{ft} improves upon this configuration, and \gls{ft} leads to a pronounced reduction in species performance.

\gls{birdmae} responds most favourably to \gls{ap}, which yields its strongest results for both species and call-type recognition. Its token-level representation appears to benefit from learning how to selectively aggregate informative regions of the input, without modifying the pretrained encoder. The weaker \gls{ft} results, particularly for species recognition, indicate that adapting the aggregation mechanism is more effective than updating the full backbone in this setting.

For \gls{eat}, \gls{lp} performs best for species recognition, whereas \gls{ft} produces a marked improvement for call types. This contrast suggests that the frozen representation retains transferable information for species identification, but is less directly aligned with the acoustic distinctions required for call-type classification. End-to-end adaptation may therefore be necessary to refine the representation towards call-type-specific cues.

A similar task-dependent pattern is observed for \gls{protoclr}. \gls{ap} is most effective for species recognition, indicating that selective aggregation improves the use of its frozen representation, while \gls{ft} performs best for call types. This may indicate that call-type distinctions require deeper adaptation of the contrastively pretrained features than species recognition.

Overall, ConvNeXt$_{BS}$ transfers most effectively through \gls{lp}, and \gls{birdmae} through \gls{ap}. For \gls{eat} and \gls{protoclr}, the preferred adaptation regime depends more strongly on the target task.

\subsection{Effect of Adaptive Loss Weighting}

The effect of adaptive loss weighting varies across adaptation regimes and tasks. Under \gls{lp}, differences between Naive, Uncertainty, and \gls{dwa} are generally modest, indicating that loss reweighting has limited influence when only the classification heads are trainable.

Under \gls{ap}, uncertainty weighting consistently provides the strongest species performance across all backbones. Its effect on call-type recognition is less uniform, with Naive remaining preferable for most backbones and Uncertainty performing best for \gls{birdmae}.

Under \gls{ft}, \gls{dwa} achieves the best species performance for most backbones, while Uncertainty is marginally stronger for \gls{protoclr}. Adaptive weighting also performs best for call-type recognition in most \gls{ft} configurations, although ConvNeXt$_{BS}$ continues to favour Naive.

Overall, adaptive weighting provides a more consistent benefit for species recognition than for call-type recognition. For species, Uncertainty performs best under \gls{ap}, while \gls{dwa} is preferred under \gls{ft} for most backbones. For call-type recognition, the differences are smaller and less systematic, with both adaptive strategies and the manually weighted Naive baseline performing best in different configurations. The preferred weighting strategy therefore depends on the task, backbone, and adaptation regime.

\subsection{Gradient Normalisation under Full Fine-Tuning}
Table~\ref{tab:finetune_loss_strategies} compares \gls{gradnorm} with the other loss-weighting strategies under \gls{ft}. \gls{gradnorm} does not improve species recognition for any backbone. Its effect on call-type recognition is more favourable but remains backbone-dependent: it performs best for ConvNeXt$_{BS}$ and \gls{protoclr}, whereas \gls{dwa} remains preferable for \gls{eat} and \gls{birdmae}. \gls{gradnorm} therefore does not provide a consistent advantage across both tasks.

The observed pattern suggests that gradient-based balancing shifts optimisation towards the call-type objective for some backbones, improving call-type recognition at the expense of species performance. Moreover, \gls{gradnorm} requires additional task-specific gradient computations with respect to the shared trainable backbone parameters, resulting in substantially higher computational and memory costs than the other weighting strategies. Given this overhead, its limited and task-dependent improvements do not represent a favourable performance-cost trade-off in the evaluated setting.

\subsection{Task-Specific Behaviour}

The results reveal different optimisation requirements for the two prediction tasks. Species recognition is more sensitive to preserving the pretrained representation and benefits consistently from adaptive loss balancing. Call-type recognition is more backbone dependent: it can benefit substantially from \gls{ft}, as observed for \gls{eat} and \gls{protoclr}, but remains strongest with frozen-backbone adaptation for ConvNeXt$_{BS}$ and \gls{birdmae}.

Peak and average performance also lead to slightly different conclusions. \gls{birdmae} with \gls{ap} and uncertainty weighting achieves the highest individual results for both tasks, whereas ConvNeXt$_{BS}$ with \gls{lp} provides the highest mean species score across weighting strategies. Reporting the two prediction heads separately is therefore important, as aggregated multi-task performance could obscure differences in task sensitivity and configuration robustness.

\begin{table}[!htb]
    \centering
    \footnotesize
    \caption{Class-masked \gls{cmap} results on the WiWa dataset. Colours indicate the adaptation regimes (\gls{lp}, \gls{ap}, and \gls{ft}), and Score denotes the mean across loss-weighting strategies. Underlining marks the best loss-weighting result within each regime, while boldface indicates the best adaptation regime by Score for each backbone-task pair. The symbols $^{*}$ and $^{**}$ denote the best and second-best backbone Scores for each task, respectively.}

    %\resizebox{0.7\textwidth}{!}{%
    \input{Tables/Table3_main_result}

    \label{tab:main-results}
    %}
\end{table}

\begin{table}[!htb]
    \centering
    \footnotesize
    \caption{Class-masked \gls{cmap} results under \gls{ft} for different loss-weighting strategies. Underlining marks the best loss-weighting strategy for each backbone and task, while boldface indicates the best backbone for each loss-weighting strategy and task.}

    \small
    \setlength{\tabcolsep}{4pt}
    \renewcommand{\arraystretch}{1.15}
        \resizebox{\textwidth}{!}{%

\input{Tables/Table4_result_gradnorm}

        \label{tab:finetune_loss_strategies}
        }
    \end{table}

\subsection{Limitations}

The present study focuses on a controlled comparison involving one benchmark and four pretrained backbones. As the models differ in their input representations, sampling rates, and input durations, the results reflect the performance of their complete processing pipelines rather than architectural differences alone. All configurations were evaluated over three seeds, and the reported means provide a consistent basis for comparison, although very small differences should be interpreted with appropriate caution.

The combined single-task and factorised multi-task formulations operate on different label spaces. Their results therefore provide complementary evidence about the two modelling approaches rather than a strictly equivalent metric comparison. \gls{gradnorm} is evaluated under \gls{ft}, where shared backbone parameters are trainable, and uses a smaller batch size to accommodate its additional gradient computations.

The current class-masking procedure is applied consistently across all experiments and provides a conservative evaluation. However, retaining the original output space may allow classes without positive test examples to influence the macro-average, particularly for call-type recognition, where such cases are assigned zero. Future evaluations could refine this procedure by excluding undefined classes directly from the final average.

\section{Conclusions}
\label{sec:conclusion}

This work investigated adaptive loss balancing for joint bird species and call-type classification on the long-tailed WiWa benchmark. By comparing a fixed weighted objective with uncertainty weighting, \gls{dwa}, and \gls{gradnorm} across four pretrained backbones and three adaptation regimes, the experiments show that the effectiveness of multi-task optimisation depends strongly on both the backbone representation and the chosen adaptation regime.

The factorised multi-task formulation shows its clearest and most consistent advantage for call-type recognition, whereas its effect on species recognition depends more on the adaptation regime. \gls{ft} is therefore not consistently the best choice. ConvNeXt$_{BS}$ achieves the highest mean species performance with \gls{lp}, whereas \gls{birdmae} with \gls{ap} provides the highest mean call-type performance. Moreover, \gls{birdmae} with \gls{ap} and uncertainty weighting achieves the highest individual result for both tasks. These findings indicate that preserving pretrained representations can be more effective and computationally efficient than updating the complete encoder.

Adaptive weighting improves most configurations, but no strategy dominates universally. For species recognition, uncertainty weighting is particularly effective under \gls{ap}, whereas \gls{dwa} is generally more competitive under \gls{ft}. The call-type results are less systematic and depend more strongly on the backbone and adaptation regime. \gls{gradnorm} improves call-type recognition for selected backbones, but its higher computational cost and consistently weaker species performance limit its practical advantage.

Overall, loss balancing should be selected jointly with the backbone, adaptation regime, and target task rather than treated as an independent optimisation choice. Future work should evaluate the generalisability of these findings on additional bioacoustic datasets and refine evaluation for sparse and partially overlapping label spaces by excluding classes without positive evaluation samples from the macro-average rather than assigning them zero scores.

\begin{credits}
\subsubsection{\ackname} This research was funded by the German Federal Ministry for the Environment, Nature Conservation, Nuclear Safety and Consumer Protection (BMUV) through the project “DeepBirdDetect -– Automatic Bird Detection of Endangered Species Using Deep Neural Networks” (FKZ 67KI31040C). The study makes use of data collected within the project “Optimierung des Planungs- und Genehmigungsverfahrens für Windenergieanlagen im Wald im Hinblick auf artenschutzrechtliche Belange (Avifauna)” (FKZ 3517 86 0400), funded by the German Federal Agency for Nature Conservation (BfN) with financial support from the BMUV.

\subsubsection{\discintname}
The authors have no competing interests to declare that are relevant to the content of this article.

\end{credits}

% ---- Bibliography ----

\bibliographystyle{splncs04}
\bibliography{Adaptive_loss_citation}

\end{document}

%% file: Tables/Table1_Hyperparameter.tex
\begin{tabular}{lccc}
\toprule
\textbf{Hyperparameter} & \textbf{LP} & \textbf{AP} & \textbf{FT} \\
\midrule
Learning rate           & $5\times10^{-3}$ & $1.3\times10^{-3}$ & $1\times10^{-4}$ \\
LR schedule             & Cosine           & Cosine             & Cosine           \\
Warmup ratio            & 5\%              & 5\%                & 5\%              \\
Max epochs              & 15               & 20                 & 20               \\
Batch size              & 128              & 128                & 128 (8$^\dagger$) \\
Gradient clipping       & 0.5              & 0.5                & 0.5              \\
Early stopping patience & 3                & 5                  & 5                \\
Min $\Delta$ (\gls{stl}) & $1\times10^{-4}$ & $1\times10^{-4}$   & $1\times10^{-4}$ \\
Min $\Delta$ (\gls{mtl}) & $2\times10^{-4}$ & $2\times10^{-4}$   & $2\times10^{-4}$ \\
\bottomrule
\end{tabular}

%% file: Tables/Table2_lossstrategy_applicability.tex
\begin{tabular}{lccc}
\toprule
\textbf{Strategy} & \textbf{LP} & \textbf{AP} & \textbf{FT} \\
\midrule
Fixed Weighting ($\lambda_s{=}1.0,\;\lambda_c{=}17.0$) & \checkmark & \checkmark & \checkmark \\
Uncertainty Weighting                       & \checkmark & \checkmark & \checkmark \\
Dynamic Weight Averaging ($T{=}2.0$)        & \checkmark & \checkmark & \checkmark \\
GradNorm ($\alpha{=}1.5$)                   & --         & --         & \checkmark \\
\bottomrule
\end{tabular}

%% file: Tables/Table3_main_result.tex
\renewcommand{\arraystretch}{1.5}
\setlength{\tabcolsep}{2pt}
\small

\resizebox{\textwidth}{!}{%
\begin{tabular}{%
  >{\centering\arraybackslash}p{1.9cm}
  p{1.5cm}
  @{\vrule width 1pt}
  >{\centering\arraybackslash}p{1.5cm}
  @{\vrule width 1pt}
  >{\centering\arraybackslash}p{1.2cm}
  >{\centering\arraybackslash}p{1.2cm}
  >{\centering\arraybackslash}p{1.2cm}
  >{\centering\arraybackslash}p{1.4cm}
  @{\vrule width 1pt}
  >{\centering\arraybackslash}p{1.2cm}
  >{\centering\arraybackslash}p{1.2cm}
  >{\centering\arraybackslash}p{1.2cm}
  >{\centering\arraybackslash}p{1.4cm}
}
    \toprule
    \multicolumn{2}{c}{} &
      \multicolumn{1}{c}{Single-Task} &
      \multicolumn{8}{c}{Multi-Task} \\
    
    \addlinespace[2pt]
    \arrayrulecolor{black}\cline{3-11}
    \addlinespace[2pt]
    
    \rowcolor{gray!50}
    \multicolumn{2}{c}{Setting} &
      \multicolumn{1}{c}{Combined} &
      \multicolumn{4}{c}{Species} &
      \multicolumn{4}{c}{Call Type} \\
    
    \addlinespace[2pt]
    \arrayrulecolor{black}\cline{1-11}
    \addlinespace[2pt]
    
    \rowcolor{gray!15}
    \multicolumn{2}{c}{Loss Strategy} &
      \multicolumn{1}{c}{} &
      Naive &
      Unc. &
      DWA &
      Score &
      Naive &
      Unc. &
      DWA & 
      Score \\
    
    \midrule

    \multicolumn{11}{p{5cm}}{\textit{Bioacoustic models}} \vspace{0.5mm} \\

    % -------- ConvNeXt_BS --------
    \multirow{3}{*}{%
      \rotatebox[origin=c]{360}{%
        \renewcommand{\arraystretch}{1}%
        \begin{tabular}{@{}c@{}}
          ConvNext$_{BS}$
        \end{tabular}}} &
      \cellcolor{gray!10} Linear &
      $0.3481$ &
      $0.4470$ &
      $0.4472$ &
      $\uline{0.4505}$ &
      \textbf{\heatgreen{0.4482}}\textsuperscript{$\star$} &
      $0.499$ &
      $\uline{0.5007}$ &
      $0.5003$ & 
      \textbf{\heatgreen{0.50}}\textsuperscript{$\star\star$} \\

      \arrayrulecolor{lightgray}\cline{2-11}

    & Attentive &
      $0.3108$ &
      $0.3802$ &
      $\uline{0.4115}$ &
      $0.3926$ &
      \heatorange{0.3947} &
      $\uline{0.4713}$ &
      $0.4539$ &
      $0.4631$ &
      \heatorange{0.4627} \\

      \arrayrulecolor{lightgray}\cline{2-11}

    & \cellcolor{gray!30} Finetune &
      $\textbf{0.3852}$ &
      $0.3153$ &
      $0.3272$ &
      $\uline{0.3322}$ &
      \heatblue{0.3249} &
      $\uline{0.4902}$ &
      $0.4652$ &
      $0.4736$ &
      \heatblue{0.4763} \\

    \arrayrulecolor{black}\hline

    % -------- EAT --------
    \multirow{3}{*}{\rotatebox[origin=c]{360}{EAT}} &
      \cellcolor{gray!10} Linear &
      $0.1880$ &
      $0.2886$ &
      $\uline{0.2940}$ &
      $0.2929$ &
      \textbf{\heatgreen{0.2918}} &
      $0.3473$ &
      $\uline{0.3537}$ &
      $0.3524$ &
      \heatgreen{0.3511} \\

      \arrayrulecolor{lightgray}\cline{2-11}

    & Attentive &
      $0.1695$ &
      $0.2851$ &
      $\uline{0.2890}$ &
      $0.2797$ &
      \heatorange{0.2846} &
      $\uline{0.3453}$ &
      $0.3415$ &
      $0.3350$ &
      \heatorange{0.3406} \\

      \arrayrulecolor{lightgray}\cline{2-11}

    & \cellcolor{gray!30} Finetune &
      $\textbf{0.2087}$ &
      $0.2053$ &
      $0.2205$ &
      $\uline{0.2223}$ &
      \heatblue{0.2160} &
      $0.4484$ &
      $0.4481$ &
      $\uline{0.4495}$ &
      \textbf{\heatblue{0.4486}}\\

    \arrayrulecolor{black}\hline

    % -------- BirdMAE --------
    \multirow{3}{*}{%
      \rotatebox[origin=c]{360}{%
        \renewcommand{\arraystretch}{1}%
        \begin{tabular}{@{}c@{}}
          BirdMAE
        \end{tabular}}} &
      \cellcolor{gray!10} Linear &
      $0.2390$ &
      $0.3555$ &
      $\uline{0.3616}$ &
      $0.3521$ &
      \heatgreen{0.3564} &
      $\uline{0.4850}$ &
      $0.4845$ &
      $0.4830$ &
      \heatgreen{0.4841}\\

      \arrayrulecolor{lightgray}\cline{2-11}

    & Attentive &
      $\textbf{0.4111}$ &
      $0.4155$ &
      $\uline{0.4641}$ &
      $0.4493$ &
      \textbf{\heatorange{0.4429}}\textsuperscript{$\star\star$} &
      $0.5199$ &
      $\uline{0.5359}$ &
      $0.5342$ &
      \textbf{\heatorange{0.53}}\textsuperscript{$\star$} \\

      \arrayrulecolor{lightgray}\cline{2-11}

    & \cellcolor{gray!30} Finetune &
      $0.3351$ &
      $0.2445$ &
      $0.2769$ &
      $\uline{0.2906}$ &
      \heatblue{0.2706} &
      $0.4924$ &
      $0.4916$ &
      $\uline{0.4965}$ &
      \heatblue{0.4935}\\

    \arrayrulecolor{black}\hline

    % -------- ProtoCLR --------
    \multirow{3}{*}{%
        \renewcommand{\arraystretch}{1}%
        \begin{tabular}{@{}c@{}}
          ProtoCLR
        \end{tabular}} &
      \cellcolor{gray!10} Linear &
      $0.2048$ &
      $0.3066$ &
      $0.3075$ &
      $\uline{0.3091}$ &
      \heatgreen{0.3077} &
      $0.3688$ &
      $0.3672$ &
      $\uline{0.3708}$ &
      \heatgreen{0.3689}\\

      \arrayrulecolor{lightgray}\cline{2-11}

    & Attentive &
      $0.2446$ &
      $0.3137$ &
      $\uline{0.3394}$ &
      $0.3296$ &
      \textbf{\heatorange{0.3275}} &
      $\uline{0.3993}$ &
      $0.3975$ &
      $0.3975$ &
      \heatorange{0.3981} \\

      \arrayrulecolor{lightgray}\cline{2-11}

    & \cellcolor{gray!30} Finetune &
      $\textbf{0.2687}$ &
      $0.2250$ &
      $\uline{0.2535}$ &
      $0.2526$ &
      \heatblue{0.2437} &
      $0.4477$ &
      $\uline{0.4622}$ &
      $0.4567$ &
       \textbf{\heatblue{0.4555}}\\

    \arrayrulecolor{black}
    \bottomrule

\end{tabular}%
}

%% file: Tables/Table4_result_gradnorm.tex
\begin{tabular}{lcccccccc}
\toprule
\multirow{2}{*}{\textbf{Backbone}} 
& \multicolumn{4}{c}{\textbf{Species}} 
& \multicolumn{4}{c}{\textbf{Call Type}} \\
\cmidrule(lr){2-5} \cmidrule(lr){6-9}
& \textbf{Naive} 
& \textbf{Unc.} 
& \textbf{DWA} 
& \textbf{GradNorm}
& \textbf{Naive} 
& \textbf{Unc.} 
& \textbf{DWA} 
& \textbf{GradNorm} \\
\midrule

ConvNeXt$_{BS}$ 
& $\textbf{0.3095}$ & $\textbf{0.3272}$ & $\uline{\textbf{0.3322}}$ & $\textbf{0.2864}$ 
& $\textbf{0.4946}$ & $0.4652$ & $0.4736$ & $\uline{\textbf{0.5038}}$ \\

EAT 
& $0.2027$ & $0.2205$ & $\uline{0.2223}$ & $0.1204$ 
& $0.4416$ & $0.4481$ & $\uline{0.4495}$ & $0.3707$ \\

BirdMAE 
& $0.2339$ & $0.2769$ & $\uline{0.2906}$ & $0.2014$ 
& $0.4939$ & $\textbf{0.4916}$ & $\uline{\textbf{0.4965}}$ & $0.4648$ \\

ProtoCLR 
& $0.2315$ & $\uline{0.2535}$ & $0.2526$ & $0.2311$ 
& $0.4412$ & $0.4622$ & $0.4567$ & $\uline{0.4647}$ \\

\bottomrule
\end{tabular}